\documentclass[a4paper]{article}

\usepackage{INTERSPEECH2021}
\usepackage{cite, subcaption}

\title{Extremely Low Footprint End-to-End ASR System for Smart Device}
\name{Zhifu Gao$^{1*}$, Yiwu Yao$^{2*} \thanks{* Equal contribution.}$,  Shiliang Zhang$^{1}$, Jun Yang$^2$, Ming Lei$^1$, Ian McLoughlin$^3$}
\address{
	$^1$Speech Lab, Alibaba Group, China \\
	$^2$Platform of A. I., Alibaba Group, China \\
	$^3$ICT Cluster, Singapore Institute of Technology, Singapore}
\email{\{zhifu.gzf, yiwu.yyw, sly.zsl, muzhuo.yj, lm86501\}@alibaba-inc.com, ian.mcloughlin@singaporetech.edu.sg}

\begin{document}
	
	\maketitle
	\begin{abstract}
		Recently, end-to-end (E2E) speech recognition has become popular, since it can integrate the acoustic, pronunciation and language models into a single neural network, which outperforms conventional models.
		Among E2E approaches, attention-based models, $e.g.$ Transformer, have emerged as being superior.
		Such models have opened the door to deployment of ASR on smart devices, however they still suffer from requiring a large number of model parameters.
		\
		We propose an extremely low footprint E2E ASR system for smart devices, to achieve the goal of satisfying resource constraints without sacrificing recognition accuracy.
		We design cross-layer weight sharing to improve parameter efficiency and further exploit model compression methods including sparsification and quantization, to reduce memory storage and boost decoding efficiency.
		\
		We evaluate our approaches on the public AISHELL-1 and AISHELL-2 benchmarks. 
		On the AISHELL-2 task, the proposed method achieves more than 10$\times$ compression (model size reduces from 248 to 24MB), at the cost of only minor performance loss (CER reduces from 6.49\% to 6.92\%).
		
	\end{abstract}
	
	\section{Introduction}
	
	There is growing interest in building automatic speech recognition (ASR) systems on smart devices to satisfy  privacy, security and network bandwidth constraints. Recent advances in end-to-end ASR (E2E-ASR) means such systems are now strong candidates for such deployments.
	\
	Compared to conventional hybrid ASR systems, E2E-ASR systems fold the acoustic, language and pronunciation models into a single sequence-to-sequence (seq2seq) model.
	\
	Currently, there exist three popular E2E approaches, namely connectionist temporal classification (CTC)~\cite{graves2006connectionist}, recurrent neural network transducers (RNN-T)~\cite{graves2012sequence}, and attention based encoder-decoders (AED)~\cite{bahdanau2014neural,chorowski2015attention,chan2016listen}. 
	Unlike CTC-based models, RNN-T and AED have no independence assumption, and can achieve state-of-the-art performance without an external language model, making them more suitable for on-device deployment.
	
	Typical AED models such as LAS~\cite{chan2016listen} and Transformer~\cite{vaswani2017attention}, consist of an encoder and a decoder. 
	The encoder transforms raw acoustic features into a high-level representation, while the decoder predicts output symbols in an auto-regressive manner. 
	Transformer-based models have dominated seq2seq modeling in the ASR field, due to their superior recognition accuracy~\cite{sainath2018improving,raffel2017online,chiu2017monotonic,fan2018online,miao2019online,moritz2019triggered,moritz2020streaming,zhang2020streaming}.
	\
	However, the large number of model parameters these ASR models require has become the main challenge for deployment in resource constrained scenarios, where both memory and computation resources are limited.
	\
	Recently, researchers have made more effort to optimize ASR systems for smart device.
	\
	For example, knowledge distillation~\cite{pang2018compression}, 
	low-rank factorization~\cite{winata2020lightweight}, model sparsification~\cite{wu2020dynamic} and network architecture search~\cite{David2019icml}, etc.
	\
	Such efforts have made some progress, nevertheless widespread deployment of ASR on smart devices remains a challenge; both in terms of memory and computational resource constraints. 	

	\begin{figure}[t]
		\centering
		\includegraphics[width=0.75\linewidth]{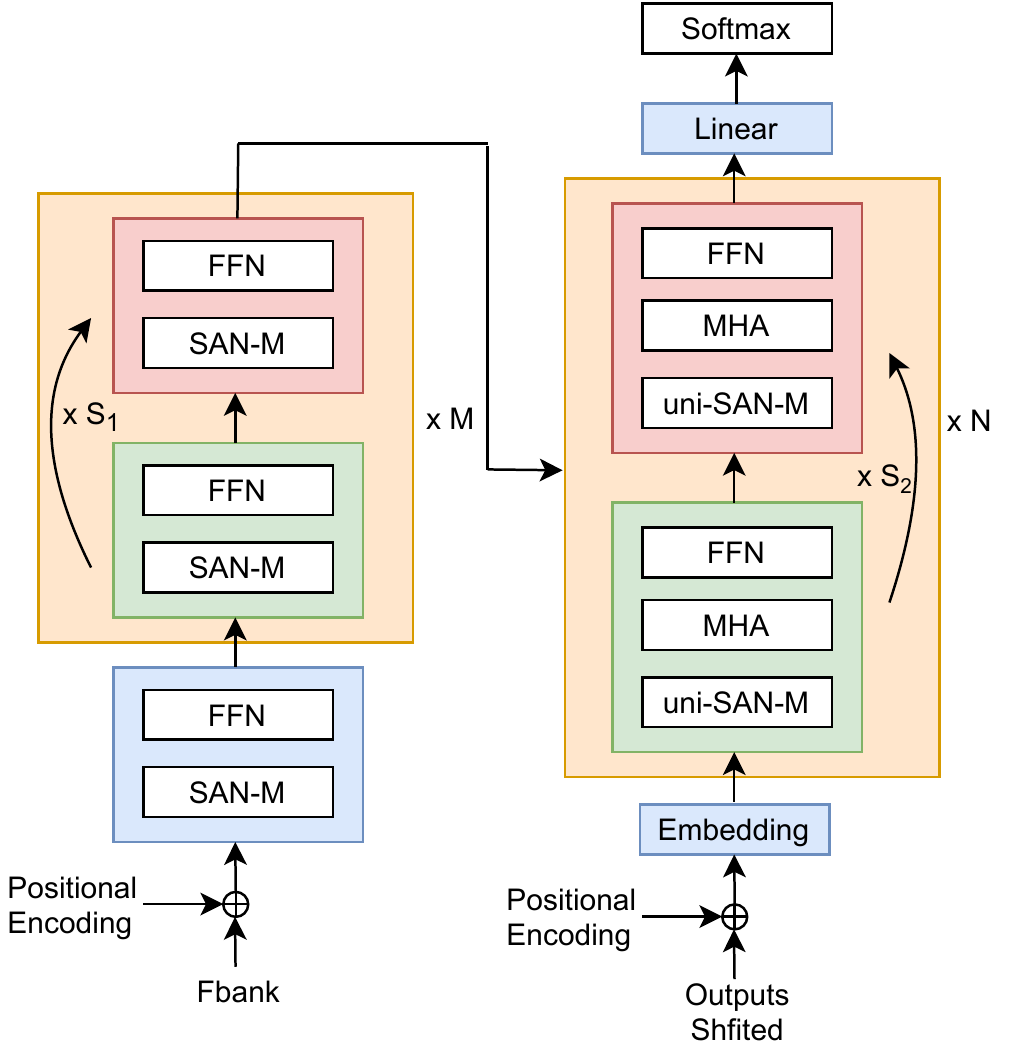}
		\caption{Illustration of the encoder-decoder arrangement with weight sharing. Colour denotes different sub-layers. The blue and green sub-layers have their own parameters while red sub-layers share the green sub-layer weights.}
		\label{fig:dfsmn}
		\vspace{-3mm}
	\end{figure}

	In this work, we propose an extremely low footprint E2E ASR system for smart devices, to better trade-off computation resources and recognition accuracy.
	\
	Firstly, motivated by ALBERT in neural language processing~\cite{lan2019albert},
	we adopt cross-layer weight sharing to improve parameter efficiency.
	This prevents the number of model parameters from growing with the depth of the network, without seriously impacting performance.
	\
	In addition to weight sharing for the ASR models, we incorporate model sparsification~\cite{Pavlo2019cvpr, hansong2016iclr, Ren2019asplos, Namhoon2019iclr, Yuhsiang2020a100} and post-training quantization (PTQ)~\cite{Steven2020iclr, Sangil2019cvpr, TRT20, Cong2017admm, Markus2019iccv, Yaohui2020cvpr} methods for extreme model compression. 
	\
	
	As shown in Fig.~\ref{fig:dfsmn}, the architecture of the proposed weight-shared ASR model contains an encoder and decoder, which adopts memory equipped self-attention (SAN-M)~\cite{gao2020san} and a feed-forward network (FFN) as the basic sub-layer. 
	SAN-M and FFN are designed to be shared across different layers, to improve the parameter efficiency.
	The ASR model is pre-trained before fine-tuning with model sparsification. Then during sparse training, the weight importance, with Taylor expansion~\cite{Pavlo2019cvpr}, is utilized for sparse constraints,  where the redundant weights of the model are zeroed out to reduce model storage size. After sparsification, the sparse model can be further quantized with negligible accuracy loss at the post-training stage, with the selected combination of a refined KL algorithm~\cite{TRT20}, improved ADMM (Alternating Direction Method of Multipliers)~\cite{Cong2017admm} or label-free automatic mixed precision (AMP)~\cite{Yaohui2020cvpr} quantization. Quantization converts the ASR model to a lower-bit representation for integer storage and significant computation acceleration.

	We report extensive experiments on the public AISHELL-1 and AISHELL-2 benchmarks.
	Experimental results show that weight-sharing improves parameter-efficiency without sacrificing model accuracy.
	When combining weight sharing with model compression, we obtain an extremely low footprint model.
	Specially, the proposed method achieves 10$\times$ compression (model size from 248MB to 24MB) at the cost of only a small performance decay (CER from 6.49\% to 6.92\%) on AISHELL-2.
	
	\section{Methods}
	
\begin{figure}[t]
		\centering
		\includegraphics[width=\linewidth]{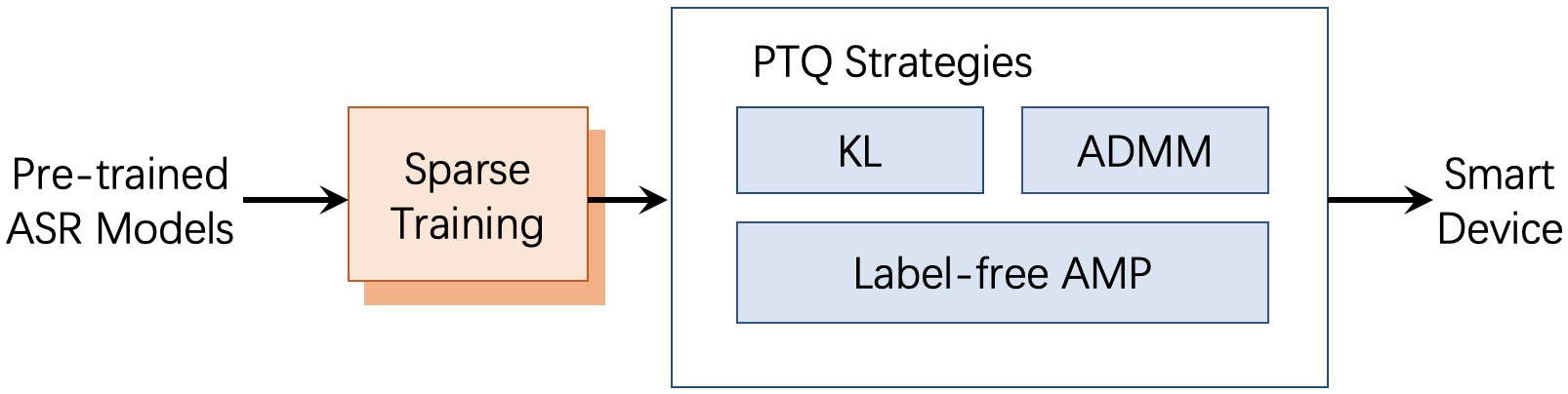}
		\caption{Overall compression framework for ASR models.}
		\label{fig:pipeline}
 		\vspace{-3mm}
	\end{figure}

	In this section, we will describe the details of the weight sharing and model compression methods we use to achieve extremely low footprint E2E ASR. The overall compression framework is depicted in Fig.~\ref{fig:pipeline}.

	\subsection{Cross-layer Weight Sharing}
	\label{sec:weight_sharing}

	As depicted in Fig.~\ref{fig:dfsmn}, the backbone of the proposed model architecture is similar to Transformer~\cite{vaswani2017attention}, while the self-attention is replaced with SAN-M~\cite{gao2020san} for both encoder and decoder. We denote the combination of SAN-M and FFN, or the combination of SAN-M, MHA and FFN as the basic sub-layer, which are tagged with the same colour in Fig.~\ref{fig:dfsmn}.
	
	\
    The encoder, consisting of a stem sub-layer and $N$ blocks made of sub-layers, maps the input sequence $\mathbf{X}$ to a sequence of hidden representations $\mathbf{Z}$.
    Each block within the encoder contains $S_{1}$ basic sub-layers which share the same weights across layers to make the block high parameter efficiency. Obviously, the total storage size of blocks in the encoder can be theoretically reduced to $1/S_{1}$ of a system without weight sharing.
    
    The decoder, meanwhile, generates one element of output sequence $\mathbf{Y}$ at each time step, by consuming representations $\mathbf{Z}$ cached from the encoder.
    \
    As an auto-regressive decoder, it consumes the previously predicted characters as additional inputs when producing the next character at each step~\cite{graves2013generating}.
    \
    The decoder consists of three components: i) the first is the embedding layer, converting the text inputs to embedding vectors;
    ii) the second includes $M$ blocks, each containing $S_{2}$ sub-layers which share weights with each other to reduce the storage size. The decoder sub-layers are composed of unidirectional SAN-M, multi-head attention (MHA) and FFN; and iii) the last component is a single feed-forward sub-layer which outputs the predicted characters.
    
    In short, the storage size of the ASR model can be significantly reduced by weight sharing across sub-layers within each building block. As noted previously, further compression is achieved by two automatic model compression methods, namely sparsification and quantization. These two model-independent compression methods are orthogonally applied.

	\subsection{Model Sparsification}
	
	According to different sparse patterns, model sparsification methods can be classified into structural ~\cite{Pavlo2019cvpr} or non-structural methods~\cite{hansong2016iclr, Ren2019asplos, Namhoon2019iclr, Yuhsiang2020a100}. Non-structural sparsification randomly zeros parameters at the fine-grained level of the weights, leading to favourable accuracy robustness with high compression ratio. Therefore, in order to maintain the recognition accuracy of ASR models with high sparsity, non-structural sparsification is adopted in this work. The Taylor method~\cite{Pavlo2019cvpr}, leveraging the second term of the Taylor expansion of the pruning cost as the importance metric, is then used to reveal the weight importance:
	
	\begin{equation}
    \label{eq:taylor}
    I=\left(w \cdot g \right)^2=\left( w \cdot \dfrac{\partial L}{\partial w}\right)^2
    \end{equation}
    
    In Eq.~\ref{eq:taylor}, the backward gradient information $g$ is applied for the implication of weight importance, thus the redundancy of model parameters can be effectively explored for sparsification. To enhance the effectiveness of sparse training with the Taylor metric, four aspects are considered:
    \begin{itemize}
    \item First, the weight importance is initialized as the weight magnitude; 
    \item During training, the weight importance is updated with the Taylor method and EMA (exponential moving average); 
    \item The compression ratio for each layer is dynamically determined with the overall ranking of weight importance; 
    \item The compression ratio for the entire model is then progressively increased during sparse training.
    \end{itemize}
    
    Thereby the weight importance is updated at the training step $t$ using the following expression ($\alpha=0.99$):
    \begin{equation}
    \label{eq:ema}
	I_{t} = \left\{
             \begin{array}{lr}
             \| w_{t} \|, & t=0 \\
             \alpha \cdot I_{t-1} + (1 - \alpha) \cdot \left(w_{t} \cdot g_{t} \right)^2, & t>0   
             \end{array} \right.
    \end{equation}
             
	\subsection{Post-training Quantization}
	
	After sparse training of the weight-shared ASR model, the quantization is applied for further compression and acceleration. Post-training quantization (PTQ) is preferred in this work for a simpler compression flow. Meanwhile, symmetric quantization is exploited to map the activations and weights into an 8-bit integer range $[-127, 127]$. The quantization function $Q(\cdot)$ for a given tensor $v$ is expressed as follows:
	
	\begin{equation}
    \label{eq:quant}
    Q(v) = round(clip(v/s, -T_s, T_s))
    \end{equation}
    
    Where $s$ is the scaling factor for quantization, and $T_s = 127$ denotes the integer range. According to Eq.~\ref{eq:quant}, quantization noise is inevitably introduced to the quantized ASR model, thanks to the influence of clipping and rounding errors. 
    
    In order to minimize the quantization noise and thus maintain recognition accuracy, several PTQ strategies are explored in this work, including a refined KL algorithm, improved ADMM and label-free automatic mixed precision (AMP) quantization.
    
    \subsubsection{Refined KL Algorithm}
    
    \begin{figure}[t]
    \centering
    \begin{minipage}[t]{0.48\linewidth}
    \centering
    \includegraphics[width=\linewidth]{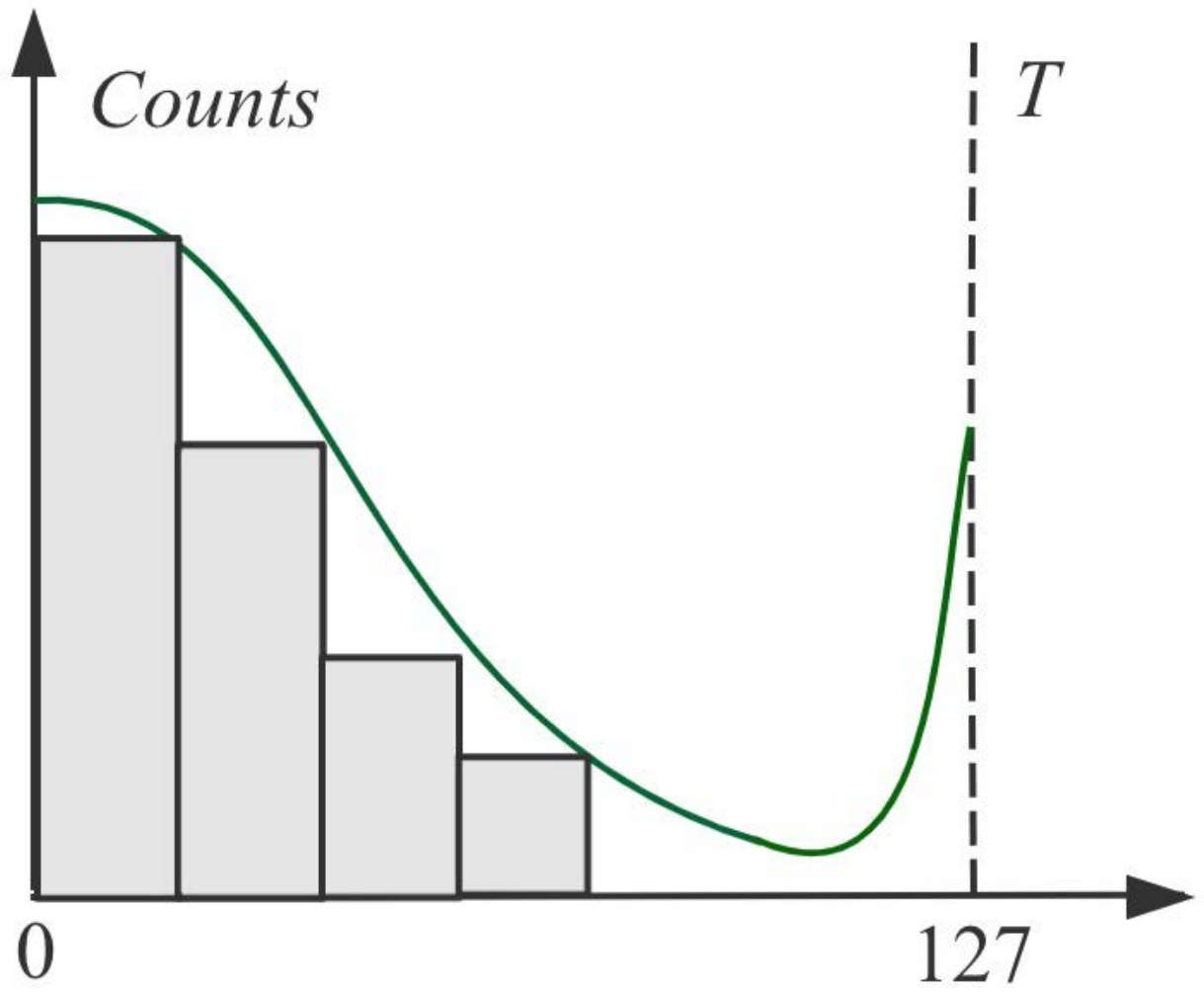}
    \subcaption{Absolute histogram}
    \label{fig:hist1}
    \end{minipage}
    \begin{minipage}[t]{0.48\linewidth}
    \centering
    \includegraphics[width=\linewidth]{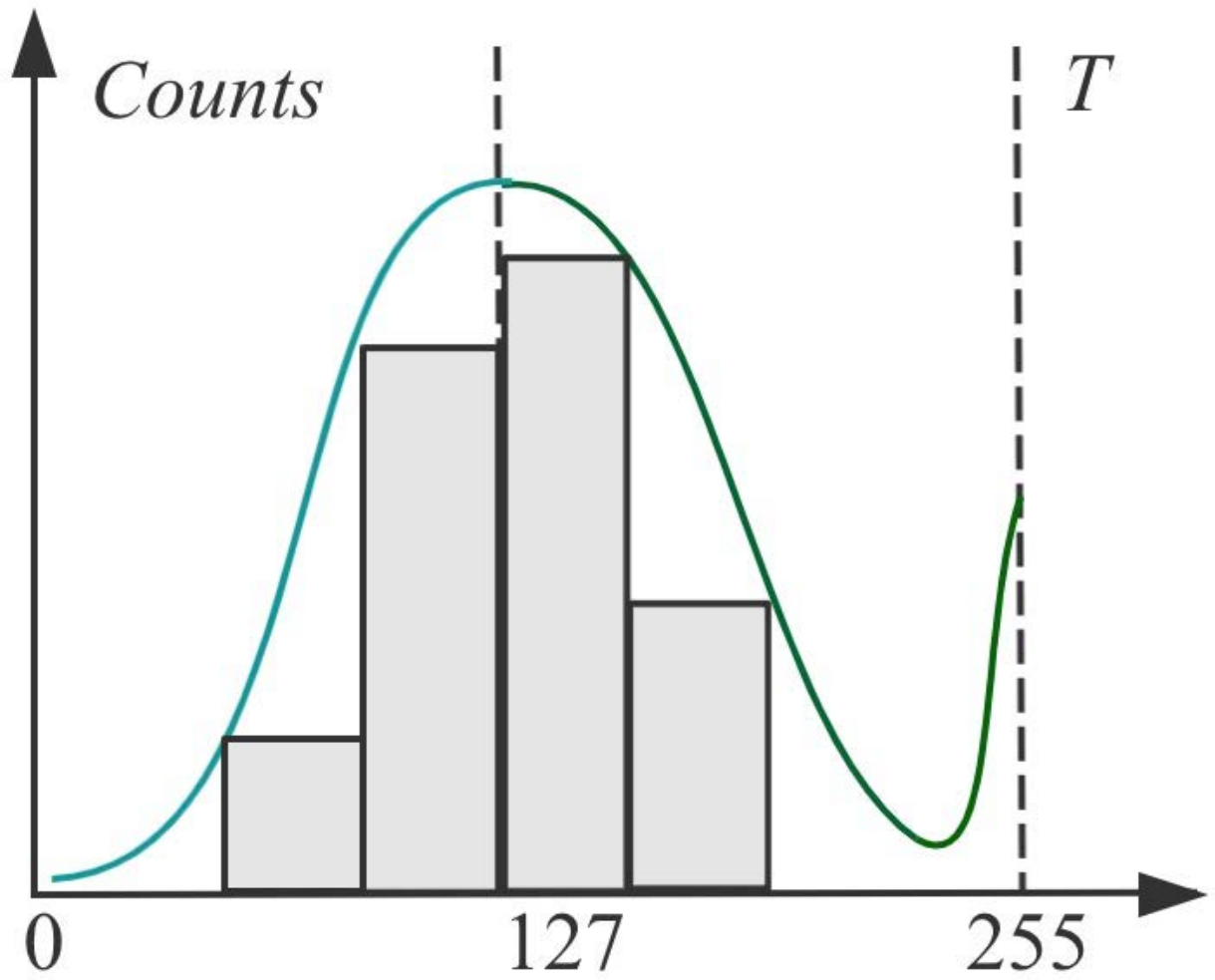}
    \subcaption{Real histogram}
    \label{fig:hist2}
    \end{minipage}
    \caption{Two histogram types for the KL algorithm: (a) absolute values are mapped into $[0, 127]$; (b) the real values are mapped into $[0, 255]$. The scaling factor is calculated as $T/127$ of the statistical threshold $T$ in the KL algorithm.}
    \label{fig:hist}
    \end{figure}
    
    The KL algorithm is used to calculate the scaling factors for the activation quantization process, as detailed in~\cite{TRT20}. As shown in Fig.~\ref{fig:hist1}, the absolute value of activations (excluding zeros) are quantized and then mapped into a histogram with an 8-bit integer range $[0, 127]$,  that is used for the quantization of ReLU activations. 
 
    Adhering to the basic principle of the KL algorithm, we refine the process by collecting the histogram over the original real activation values (excluding zeros). As shown in Fig.~\ref{fig:hist2}, the integer range is now $[0, 255]$, which broadens the histogram and captures finer statistics, meaning that the quantization resolution is nearly doubled. This improvement is beneficial in reducing quantization noise associated with activations, especially in cases where activations are transferred without ReLU, which commonly occurs in the attention modules of Transformer-based ASR models.
    
    \subsubsection{Improved ADMM}
    
    Furthermore, to reduce noise due to weight quantization, ADMM~\cite{Cong2017admm} is utilized to optimize scaling factors applied prior to the quantization, according to the following iterative formula:
    
    \begin{equation}
    \label{eq:admm}
	s_{k+1}=\frac{w \cdot E(w/s_{k})}{E(w/s_{k}) \cdot E(w/s_{k})}
    \end{equation}
    
    Where $k$ denotes the iteration step, $w$ represents a weight tensor, and $E(\cdot)$ is the expectation. In order to obtain the best scaling factor as the local optimum, we improve ADMM by adopting a winner-take-all method. The mean squared error (MSE) of the original weight and quantized weight are recorded for each iteration. Finally, the scaling factor which results in the minimum MSE, is chosen for weight quantization as follows:
    \begin{equation}
    \label{eq:si}
	s_{i}=\mathop{\arg\min}_{i}\|w - s_{i} \cdot Q(w)\|^2
    \end{equation}

    \subsubsection{Label-free Automatic Mixed Precision}
    \label{sec:AMP}
    
    After post-training quantization with the refined KL algorithm and ADMM, the quantization noise for activations and weights can be eliminated to a certain extent. However, we have found that accuracy loss may still exist due to occasional abnormal network layers within the ASR model. To overcome this issue, we apply label-free automatic mixed precision (AMP) quantization as a fallback. This overcomes the relatively higher quantization cost for those few abnormal layers.

	\begin{figure}[t]
		\centering
		\includegraphics[width=\linewidth]{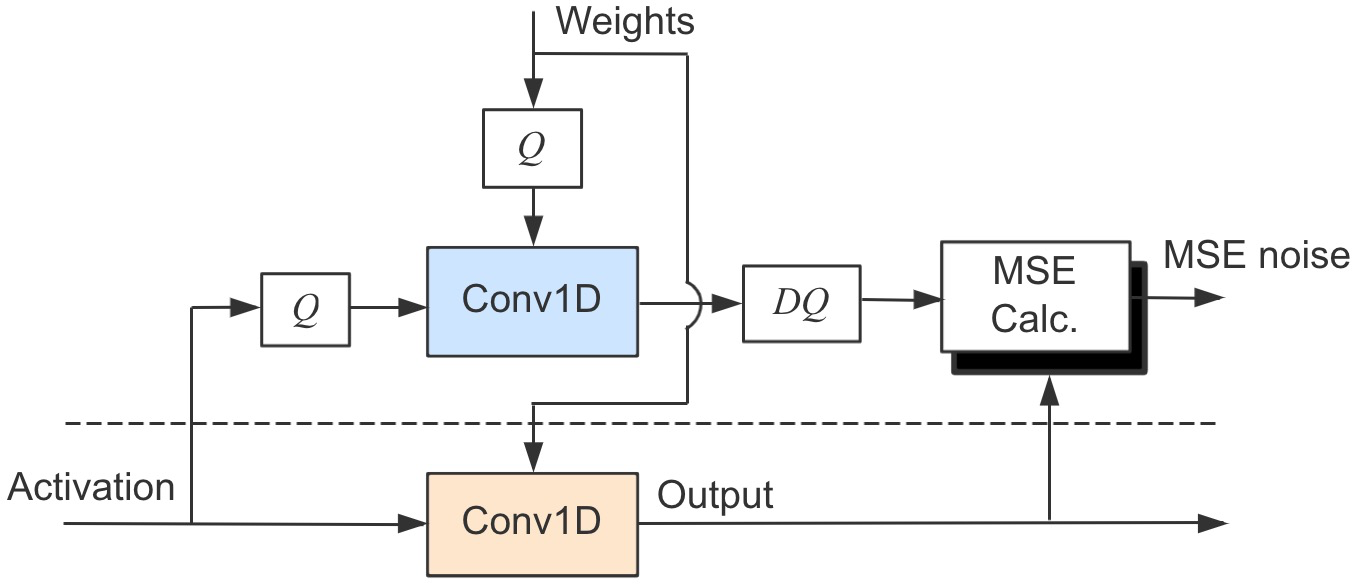}
		\caption{Sub-graph built for label-free AMP quantization.}
		\label{fig:amp}
 		\vspace{-3mm}
	\end{figure}

    As demonstrated in Fig.~\ref{fig:amp}, we build a sub-graph of fake quantization and MSE calculation in parallel to each network layer (e.g. Conv1D), to collect corresponding layer-wise MSE as the quantization cost. The calibration-set for KL algorithm, and scaling factors for quantization, are reused here for the statistics of MSE. Then the layer-wise MSE is calculated as:
    
    \begin{equation}
    \label{eq:mse}
    Cost_j=\|H_j(a,w)-DQ(H_j(Q(a), Q(w)))\|^2
    \end{equation}
    
    Where $H_j$ denotes the $j$-th layer operation (e.g. Conv1D), $DQ(\cdot)$ means the de-quantization function, and $a, w$ are the activations and weights respectively. Abnormal layers with Top-K quantization costs can, as fallback, use a float32 implementation. Since there is no need for data labels to compute the accuracy difference on an evaluation-set for the decision of mixed-precision setting, proposed AMP method is label-free. Additionally, compared with alternative label-free AMP methods~\cite{Yaohui2020cvpr} requiring a greedy search for mixed-precision settings, the time cost of our layer-wise AMP method is much lower.
	
	\section{Experiments}
	
	\subsection{Experimental Setup}
	
	We have evaluated the proposed method on two Mandarin speech recognition tasks, the 170-hour AISHELL-1~\cite{bu2017aishell} and the 1000-hour AISHELL-2 task~\cite{du2018aishell}.
    \
    For AISHELL-2, we use all the training data (1000 hours) for training, $dev\_ios$ and $test\_ios$ sets for validation and evaluation, respectively. 
    \
    Acoustic features used in all experiments are 80-dimensional log-mel filter-bank (FBK) energies computed on 25ms windows with 10ms shift.
    We stack the consecutive frames within a context window of 7 (3+1+3) to produce the 560-dimensional features and then down-sample the input frame rate to 60ms. 
    Acoustic modeling units are Chinese characters, totalling the vocabulary size of 4233 both for AISHELL-1 and AISHELL-2. 
    Models are trained with Tensorflow~\cite{abadi2016tensorflow} and 8 Nvidia GPUs.
    As to the detailed experimental setup,  we adopt LazyAdamOptimizer with $\beta_1=0.9$, $\beta_2=0.999$, and the strategy for learning rate is \emph{noam\_decay\_v2} with $d_{model}=512, warmup\_n=8000, k=4$. 
    Label smoothing and dropout regularization with a value of $0.1$ are incorporated to prevent over-fitting. 
    SpecAugment~\cite{park2019specaugment} is also used for data augmentation in all experiments.
	
	\subsection{On AISHELL-1}
	\label{sec:aishell1}
	
	In this subsection, we will evaluate the  performance of weight sharing on the AISHELL-1 task, detailed in Table~\ref{tab:weight_sharing}. Symbols $M$ and $N$ denote that the model contains $M$ encoder blocks and $N$ decoder blocks, as shown in Fig.~\ref{fig:dfsmn}. $S_{1} / S_{2}$ refers to the number of weight-shared basic sub-layers.
	
	\
The top half of Table~\ref{tab:weight_sharing} (EXP0 to 4) are baselines with different numbers of encoder blocks, each containing only one sub-layer, without weight sharing. As expected, performance increases with encoder depth (from 2 to 9), but model size also increases (from 76 to 160MB).
 The bottom half of the table  (EXP7 to 9) shows the effect when several sub-layers within each block share the same weights. Performance again improves with encoder depth (from 6 to 12), but model size is constrained to 88MB.
   Comparing EXP4 with EXP8, both of the same depth, weight sharing reduces the EXP8 model size from 160MB to 88MB at the cost of only small performance loss (CER from 6.37\% to 6.38\%). In addition, compared to only using weight sharing in the encoder, EXP5 and EXP6 results reveal that weight sharing applied to the decoder can achieve a smaller but better model.
    
    From the above experiments, we see that weight sharing prevents the model parameters from growing with the depth of the network, without seriously hurting performance.

	\begin{table}[t]
		\centering
		\caption{Evaluation of weight sharing on AISHELL-1.}
		\begin{tabular}[t]{ccccccc}
			\hline
			\bf Model & \bf M & \bf  N & $\bf S_{1}$/ $\bf S_{2}$ & \bf Size(MB) & \bf Test & \bf Dev \\
			\hline\hline
			EXP0&	1	&	3 &	-	&	76	&	9.0	&	7.86	\\
			EXP1&	2	&	3 &	-	&	88	&	7.81	&	6.87	\\
			EXP2&	4	&	3 &	-	&	112	&	6.8	&	6.09	\\
			EXP3&	6	&	3 &	-	&	136	&	6.49	&	5.8	\\
			EXP4&	8	&	3 &	-	&	160	&	6.37	&	5.77	\\
			\hline\hline
			EXP5&	1	&	3 &	8 / 1	&	76	&	6.61	&	6.03	\\
			EXP6&	1	&	1 &	8 / 3	&	52	&	6.52	&	5.8	\\
			EXP7&	2	&	3 &	2 / 1	&	88	&	6.62	&	5.92	\\
			EXP8&	2	&	3 &	3 / 1	&	88	&	6.38	&	5.79	\\
			EXP9&	2	&	3 &	4 / 1	&	88	&	6.39	&	5.73	\\
			\hline
      
		\end{tabular}
		\label{tab:weight_sharing}
	\end{table}

	\subsection{On AISHELL-2}
	
	We now evaluate the performance of weight sharing on the AISHELL-2 task, detailed in Table~\ref{tab:AISHELL2}.  $N$, $M$ and $S_{1}/S_{2}$ have the same meaning as in Sec.~\ref{sec:aishell1}. 
	
	The EXP10 baseline model has 40 encoder sub-layers and 12 decoder sub-layers.
	Compared to EXP10, the proposed model in EXP11 obtains comparable performance with 1.6$\times$ size reduction (from 248 to 151MB), thanks to  cross-layer weight sharing.
	Combining weight sharing with sparsification and quantization, more than 10$\times$ compression (model size from 248 to 24MB) is achieved, with only small performance loss (CER from 6.49\% to 6.92\%).

	\begin{table}[t]
		\centering
		\caption{Evaluation of weight sharing and model compression on AISHELL-2.}
		\begin{tabular}[t]{ccccccc}
			\hline
			\bf Model & \bf M & \bf  N & $\bf S_{1}$/ $\bf S_{2}$ & \bf Size(MB) & \bf Test & \bf Dev \\
			\hline\hline
			EXP10&	40	&	12 &	-	&	248	&	6.49	&	6.39	\\
			\hline\hline
			EXP11&	20	&	12 &	2 / 1	&	151	&	6.54	&	6.46	\\\hline
			+sparse&	20	&	12 &	2 / 1	&	72	&	6.95	&	6.71	\\
			+quant &	20	&	12 &	2 / 1	&	24	&	6.92	&	6.78	\\
			\hline
      
		\end{tabular}
		\label{tab:AISHELL2}
	\end{table}
	\begin{table}[t]
		\centering
		\caption{The effectiveness of sparsification on pre-trained ASR models with different sparsity; CR refers to compression ratio.}
		\begin{tabular}[t]{ccccc}
			\hline
			\bf Sparsity & \bf Size(MB) & \bf  CR & \bf Test & \bf Dev  \\ 
			\hline\hline
            0$\%$  & 151  & - & 6.54 & 6.46\\
            30$\%$  & 101  & 1.5$\times$ & 6.86 & 6.88 \\ 
            40$\%$  & 86  & 1.8$\times$ & 6.87 & 6.62 \\ 
            50$\%$  & 72  & 2.1$\times$ & 6.95 & 6.71 \\
			\hline
      
		\end{tabular}
		\label{tab:sparsity}
	\end{table}

   We further evaluate sparsification effectiveness. Using the pre-trained weight-shared ASR model (EXP11 in Table~\ref{tab:AISHELL2}), we conduct sparse training with a sparsity of 30$\%$, 40$\%$ and 50$\%$. Table~\ref{tab:sparsity} shows that the storage size of the ASR model effectively reduces with increasing sparsity. At 50$\%$ sparsity, we obtain 2.1$\times$ compression (model size from 151 to 72MB) with a small accuracy loss of 0.41$\%$ (CER from 6.54$\%$ to 6.95$\%$).

	\begin{table}[t]
		\centering
		\caption{Verification of PTQ strategies on the sparsified ASR model: KL means only KL algorithm is applied; KL$\dag$ means the joint usage of KL algorithm and ADMM; AMP(3) represents the AMP with three layers kept as float32; The predict layer in decoder is not quantized due to its quantization sensitivity.}
		\begin{tabular}[t]{ccccc}
			\hline
			\bf Sparsity & \bf PTQ \bf & \bf Size(MB) & \bf Test & \bf Dev \\ 
			\hline\hline
            30$\%$  & KL & 31 & 6.89 & 6.89  \\
            30$\%$  & KL$\dag$ & 31 & 6.85 & 6.86 \\ 
            40$\%$  & KL & 28 & 6.91 & 6.62 \\
            40$\%$  & KL$\dag$ & 28 & 6.89 & 6.66 \\ 
            50$\%$  & KL & 24 & 6.92 & 6.78  \\
            50$\%$  & KL$\dag$ & 24 & 6.93 & 6.76  \\ 
            50$\%$  & KL + AMP(3) & 25 & 6.90 & 6.76 \\
            50$\%$  & KL$\dag$ + AMP(3) & 25 & 6.90 & 6.73 \\
			\hline
      
		\end{tabular}
		\label{tab:ptq}
	\end{table}

    Lastly, we evaluate a PTQ strategy. After sparsification, we introduce the selected combination of PTQ strategies to quantize the ASR model with little performance decay. As illustrated in Tab.~\ref{tab:ptq}, the KL algorithm is used as the basic PTQ method, while ADMM can be adopted to minimize weight quantization noise. As a result, we can obtain an extremely compressed ASR model with a size of just 24MB and a CER of 6.92$\%$ on AISHELL-2.
    
    To further optimize the quantization, we evaluate the label-free AMP method of Section~\ref{sec:AMP} on the 50$\%$ sparsity ASR model. As listed in Table~\ref{tab:ptq}, through the fallback of three network layers with higher MSE defined in Eq.~\ref{eq:mse}, the performance of the quantized model is improved. Meanwhile, we can see that AMP based on the joint usage of KL and ADMM achieves an improvement, due to the better Pareto efficiency in the trade-off between model size and accuracy.
	
	\section{Conclusions}
	
	This work proposed an extremely low footprint E2E ASR system for smart devices, to achieve the goal of satisfying resource constraints without sacrificing recognition accuracy.
	We adopt cross-layer weight sharing to improve parameter efficiency.
	This technique prevents the model parameters from growing with the depth of the network, which reduces the number of parameters without seriously affecting performance. 
	We further utilize model compression methods including sparsification and quantization, to reduce memory storage requirements and boost decoding efficiency on smart devices.
	\
	We evaluated our approach on public AISHELL-1 and AISHELL-2 benchmarks. 
	On the AISHELL-2 task, the proposed method achieves more than 10$\times$ compression (model size from 248 to 24MB) with small accuracy loss (CER from 6.49\% to 6.92\%). 
	
	In future work, we intend to further explore more effective model compression methods such as structural sparsification and joint optimization of PTQ strategies, and evaluate the methods on more scenarios.

	\vfill\pagebreak
	
	\bibliographystyle{IEEEtran}
	
	\bibliography{mybib}
	
	
\end{document}